 \documentstyle[twocolumn,aps,amssymb,amsmath,epsf,floats]{revtex} 
\begin{document}
\draft
\font\bfmit=cmmib10
\title{Thermodynamical Study on the Heavy-Fermion Superconductor PrOs$_4$Sb$_{12}$: \\ Evidence for Field-Induced Phase Transition}
 \twocolumn[ 
 \hsize\textwidth\columnwidth\hsize\csname@twocolumnfalse\endcsname 
\author{Y. Aoki, T. Namiki, S. Ohsaki, S. R. Saha, H. Sugawara, and H. Sato} 
\address{Department of Physics, Tokyo Metropolitan University, Hachioji, Tokyo 192-0397, Japan} 

\date{Published in J. Phys. Soc. Jpn. {\bf 71} (2002) 2098.}
\maketitle

\begin{abstract}
We report measurements of low-temperature specific heat on the $4f^2$-based heavy-fermion superconductor PrOs$_4$Sb$_{12}$.
In magnetic fields above 4.5 T in the normal state, distinct anomalies are found which demonstrate the existence of a field-induced ordered phase (FIOP).
The Pr nuclear specific heat indicates an enhancement of the 4$f$ magnetic moment in the FIOP.
Utilizing a Maxwell relation, we conclude that anomalous entropy, which is expected for a single-site quadrupole Kondo model, is not concealed below 0.16 K in zero field.
We also discuss two possible interpretations of the Schottky-like anomaly at $\sim 3$ K, i.e., a crystalline-field excitation or a hybridization gap formation.
\end{abstract}
 ]
%


\narrowtext

The $f$-electron-related heavy fermion (HF) systems exhibiting superconductivity had been found only in Ce and U intermetallic compounds.
Therefore, the recent observation of the first Pr-based heavy fermion superconductivity (HFSC) in a filled skutterudite PrOs$_4$Sb$_{12}$~\cite{BauerPRB} has profound scientific significance.

The 4$f^2$ configuration of Pr ions in intermetallic compounds had been considered to be quite stable in view of no observation of strongly correlated electron behaviors until recent studies on PrInAg$_2$~\cite{Ya96} and PrFe$_4$P$_{12}$~\cite{AoPrFe4P12PRB,SaPrFe4P12PRB,SuPrFe4P12}, in which this picture breaks down.
In PrFe$_4$P$_{12}$, which is also a member of the filled skutterudites, we have shown by specific heat~\cite{AoPrFe4P12PRB}, electrical resistivity~\cite{SaPrFe4P12PRB} and de Haas-van Alphen (dHvA) effect measurements~\cite{SuPrFe4P12} that HF behaviors appear in high fields where a nonmagnetic ordered state, probably of quadrupole origin, is suppressed.
To our knowledge, PrFe$_4$P$_{12}$ is the only system in which such definitive evidence for the 4$f^2$-based Fermi-liquid HF ground state has been obtained.

Compelling evidence for the HFSC in PrOs$_4$Sb$_{12}$ was given by a large specific heat jump $\Delta C/T=0.5$ J/K$^2$ mol at $T_c=1.85$ K on a pellet of compressed powdered single crystals~\cite{BauerPRB}.
The jump is superimposed on a Schottky-like anomaly appearing at $\sim 3$ K.
Bauer {\it et al.} attributed this peak to a doublet-triplet ($\Gamma_3 - \Gamma_5$ in $O_h$-type notation) crystalline-electric-field (CEF) thermal excitation, combining with their magnetic susceptibility $\chi (T)$ and inelastic neutron scattering data.
Since the $\Gamma_3$ non-Kramers doublet ground state has quadrupole degrees of freedom, they pointed out a possibility that the HF behavior is associated with a quadrupolar Kondo effect~\cite{CoxZawad98} on the Pr-ion lattice.
In order to confirm this scenario, it is essentially important to clarify how the entropy $R \ln 2$ associated with the $\Gamma_3$ ground state is released and whether any residual entropy is hidden far below $T_c$ or not.

In this letter, we report two important findings in PrOs$_4$Sb$_{12}$ based on specific heat and magnetization measurements on high-quality single crystalline samples: (1) clear evidence for the existence of a field-induced ordered phase (FIOP) and (2) a confirmation of no anomalous entropy concealed below 0.16 K in zero field.

Single crystals of the filled skutterudite PrOs$_4$Sb$_{12}$ and the reference compound LaOs$_4$Sb$_{12}$ were grown by Sb-flux method~\cite{SuDHvA}.
The raw materials were 4N(99.99\% pure)-Pr, 4N-La, 3N-Os and 6N-Sb.
No impurity phase was detected in a powder x-ray diffraction pattern.
The lattice parameter was determined to be $a$=9.301 \AA$ $ for PrOs$_4$Sb$_{12}$ and a=9.306 \AA$ $ for LaOs$_4$Sb$_{12}$.
The observation of the dHvA oscillations in both compounds~\cite{SuDHvA} ensures high-quality of the samples. 
The electrical resistivity $\rho(T)$ for PrOs$_4$Sb$_{12}$ shows qualitatively the same behavior as reported in ref.~\cite{BauerPRB}.
No Kondo-like behavior is visible in $\rho(T)$ although we cannot conclude whether any such behavior exists or not in the small 4f-electron contribution $\rho_{4f}(T)$ estimated by subtracting $\rho(T)$ of LaOs$_4$Sb$_{12}$.
Specific heat $C(H,T)$ for $H \parallel \langle100\rangle$ was measured by a quasiadiabatic heat pulse method described in ref.~\cite{Ao98} using a dilution refrigerator equipped with an 8-T superconducting magnet. 
The temperature increment caused by each heat pulse is controlled to be $\sim$2\%.
The bulk magnetization $M(\mu_0H\le 7$\ T$,T\ge 1.9 $\ K$)$ was measured with a Quantum-Design superconducting quantum-interference device (SQUID) magnetometer.

\begin{figure}
 \centerline{\epsfxsize=7.5cm\epsfbox{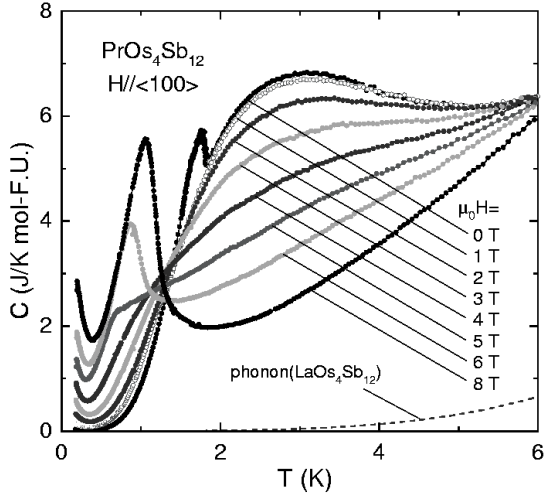}}
 \caption{Specific heat $C(T)$ of PrOs$_4$Sb$_{12}$ in different magnetic fields. The broken curve represents the phonon part $C_{ph}(T)$ determined from the $C(T)$ data of LaOs$_4$Sb$_{12}$.}
 \label{fig:CT}          
\end{figure}
Figure~\ref{fig:CT} shows the $C$-vs-$T$ data for $H\parallel\langle100\rangle$.
In zero field, the data exhibit a Schottky-like anomaly with a maximum of 6.82 J/Kmol at 3.1 K, clearly showing the existence of a low-lying excitation in the $f$-electron system.
In a $C/T$ vs $T$ plot (not shown), the peak appears at 2.1 K with a height of 2.8 J/K$^2$mol, which is 39 \% larger than the value reported in ref.~\cite{BauerPRB} for the compressed powdered single crystals, although the overall temperature dependence is similar.
In our data, a jump of $\Delta C/T=0.52$ J/K$^2$ mol associated with the SC transition is observed at $T_c=1.81$ K.
Details on the SC properties will be reported elsewhere~\cite{AokiSC}.

\begin{figure}
 \centerline{\epsfxsize=7.5cm\epsfbox{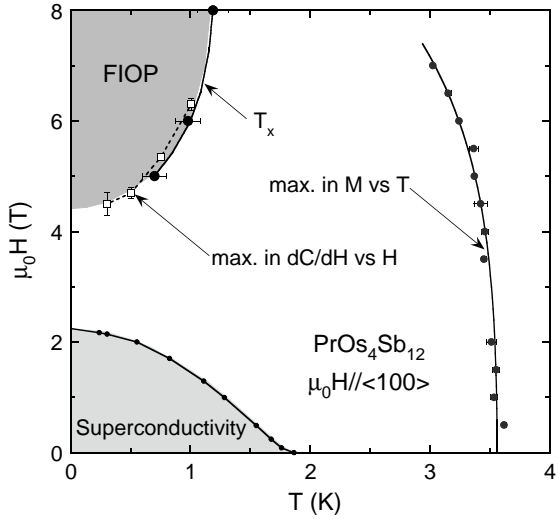}}
  \caption{
Magnetic field vs temperature phase diagram. The superconducting boundary is the data from ref.~[1]}
  \label{fig:HT}				
\end{figure}

As Fig.~\ref{fig:CT} reveals, the anomaly at $\sim 3$ K is found to be drastically suppressed with increasing magnetic field.
This field-sensitive behavior confirms that the low-lying excitation has a magnetic character. 
In 5 T, a kink appears at $\sim 0.7$ K and changes the shape into a clear $\lambda$-type peak at $T_x=0.98$ K in 6 T.
With further increasing field, the peak in the $C(T)$ curve grows and $T_x$ shifts to higher temperatures.
This is the clear thermodynamical evidence for the existence of a field-induced ordered phase (FIOP) in $T<T_x$.
The field variation of $T_x$ is plotted in a $H$-vs-$T$ phase diagram of Fig.~\ref{fig:HT}.

To estimate the phonon contribution to the specific heat, $C_{ph}$, we measured $C(T)$ of a single crystal LaOs$_4$Sb$_{12}$. 
The obtained $C_{ph}(T)$ is shown in Fig.~\ref{fig:CT}.
Negligibly small and smoothly-increasing $C_{ph}$ below 4 K strongly indicates that 4$f$ electrons play essential roles in both the Schottky-like anomaly at $\sim3$ K and the FIOP below $T_x$ in PrOs$_4$Sb$_{12}$.
\begin{figure}
 \centerline{\epsfxsize=7.5cm\epsfbox{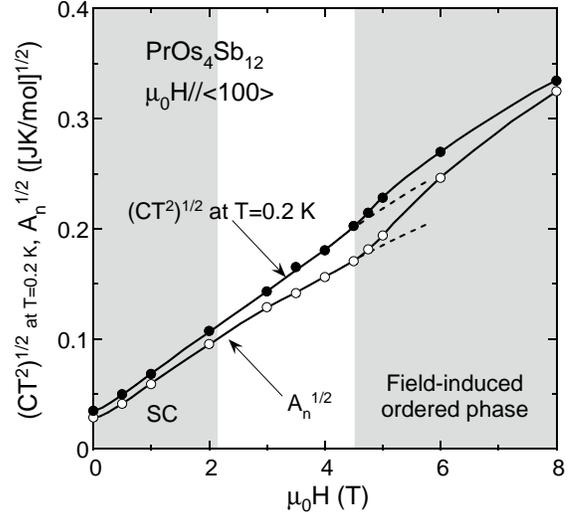}}
  \caption{Magnetic field dependence of estimated $A_n^{1/2}$ in the nuclear specific heat $C_n=A_n/T^2$. $(CT^2)^{1/2}$ at $T=0.2$ K is also shown to provide an upper bound for $A_n^{1/2}$. The lines are guides to the eye.}
\label{fig:Cn}				
\end{figure}

An upturn in $C(T)$ below 0.5 K developing with magnetic field is due to the nuclear Schottky contribution ($C_n$). 
The observed $C_n$ is mostly caused by Pr nuclei (nuclear spin $I=5/2$ for $^{141}$Pr with the natural abundance of 100\%) because of the strong intrasite hyperfine coupling between the nucleus and 4{\it f}-electrons on the same Pr ion.
This feature allows one to use $C_n$ as an on-site probe for the Pr 4$f$ magnetic moment; utilizing this, we have demonstrated that the ordered state in PrFe$_4$P$_{12}$ appearing below 6.5 K is non-magnetic in origin~\cite{AoPrFe4P12PRB}.
We analyze the $C_n$ data to obtain information on the 4$f$ magnetic moment of Pr ions in PrOs$_4$Sb$_{12}$, although the situation is complicated compared to PrFe$_4$P$_{12}$ because of a non-negligible Sb nuclear contribution.
In order to separate $C_n$ from $C$, we tentatively assume  
\begin{equation}
\label{eq:Cseparate} C(T) = A_n/T^2+\gamma T+\alpha T^n
\end{equation}  
at low temperatures.
In eq.~(\ref{eq:Cseparate}), the first term represents $C_n$ (the sum of all the nuclear contributions in PrOs$_4$Sb$_{12}$) and the other two terms represent the low-temperature excitation in $C_e(T)$.
The field dependence of $A_n^{1/2}$ obtained by fitting below 0.6 K is shown in Fig.~\ref{fig:Cn}.
To show an upper bound for $A_n^{1/2}$ based on eq.~(\ref{eq:Cseparate}), $(CT^2)^{1/2}$ at $T=0.2 $ K is also plotted.
The observed zero-field-value of $A_n^{1/2}=2.9 \times 10^{-2} $ [JK/mol]$^{1/2}$ is attributable to the Sb nuclei quadrupole contribution $2.87 \times 10^{-2} $ [JK/mol]$^{1/2}$, which is calculated from recent $^{121,123}$Sb NQR measurements~\cite{Kote2002}. 
With increasing magnetic field, $A_n^{1/2}$ increases gradually and shows a slight upward curvature around the boundary of the FIOP.
At the upper critical field $\mu_0 H_{c2}=2.2 $ T, no anomaly can be seen.
The field-dependent part in the $A_n^{1/2}$ vs $H$ curve is mostly due to the magnetic Pr nuclear contribution; the magnetic contribution from Os and Sb nuclei gives only $\sim1$ \% of the observed field dependence.
This contribution to $A_n$ can be expressed by 
\begin{equation}
\label{eq:AN} A_n^{\rm Pr} = R (A_{hf} m_{\rm Pr}/g_J)^2 I(I+1)/3,
\end{equation}  
where $R$, $A_{hf}$, $m_{\rm Pr}$ and $g_J$ are the gas constant, the magnetic dipole hyperfine coupling constant, the site-averaged magnitude of the Pr magnetic moment $g_J (\overline{|\langle J_z \rangle|^2})^{1/2}$ and the Land\'e $g$-factor, respectively.
Therefore, the field-dependent part of $A_n^{1/2}$ in Fig.~\ref{fig:Cn} reflects the $m_{\rm Pr}$ vs $H$ curve.
Using $A_{hf}=0.052 $ K, which was determined for PrFe$_4$P$_{12}$~\cite{AoPrFe4P12PRB} and is consistent with theoretical calculations~\cite{Ko61,Bl63}, the experimental value $A_n^{\rm Pr} = 0.105$ JK/mol for $\mu_0 H=8$ T leads to an estimation $m_{\rm Pr} = 1.01$ $\mu_B$/Pr in this field.
The upturn at $\sim 4.5$ T in $A_n^{1/2}(H)$ compared to a smooth extrapolation from the lower fields indicates that $m_{\rm Pr}$ (and probably also $M$) is enhanced in FIOP.

An enhancement of $M$ in the FIOP is expected independently from the Ehrenfest's theorem:
\begin{equation}
\label{eq:Ehrenfest} \Delta (\partial M/\partial T)_H=-(\Delta C/T_x) (dT_x/d(\mu_0 H)),
\end{equation}  
which should be satisfied at the second-order phase boundary of the FIOP.
Since $dT_x/dH$ is positive in the measured field range as shown in Fig.~\ref{fig:HT}, $\Delta (\partial M/\partial T)_H$ should be negative; this feature was actually observed in a recent low-temperature magnetization study~\cite{Tayama}.
For $\mu_0 H=6$ T, $\Delta (\partial M/\partial T)_H \simeq -0.04$ $\mu_B$/Pr K is calculated.
\begin{figure}
 \centerline{\epsfxsize=7.5cm\epsfbox{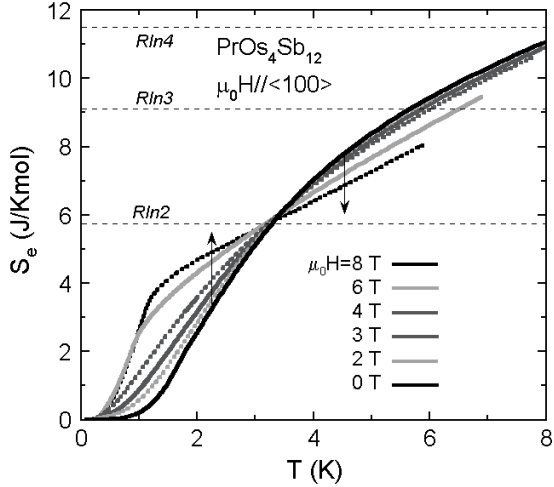}}
  \caption{Electronic part of entropy $S_{e}(T)$ calculated integrating $C_{e}(T)/T$ data. Slight adjustment has been made for the data of $H \neq 0$ so that the Maxwell relation $(\partial S/\partial (\mu_0H))_T=(\partial M/\partial T)_{\mu_0H}$ is satisfied; see text for the details.}
  \label{fig:Se}					
\end{figure} 

\begin{figure}
 \centerline{\epsfxsize=7.5cm\epsfbox{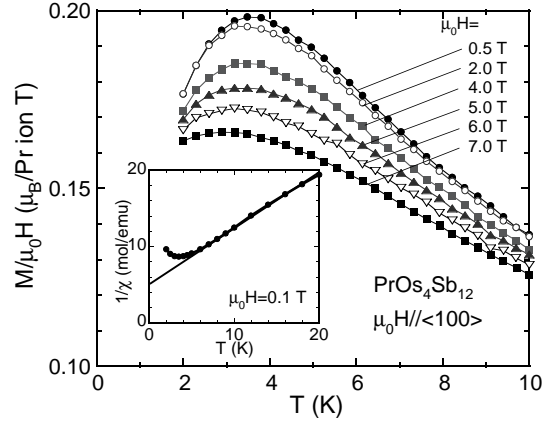}}
  \caption{Magnetization divided by applied magnetic field as a function of temperature. The inset shows the temperature dependence of inverse magnetic susceptibility.}
  \label{fig:MT}				
\end{figure}

We obtained the temperature dependence of the electronic part of entropy $S_e(T)$ by numerically integrating the data of $C_e/T \equiv (C - C_n - C_{ph})/T$ vs $T$.
Since our measurements are made above 0.16 K, only $\Delta S_e(T) \equiv S_e(T)-S_e(0.16 $ K$)$ can be obtained from the present study.
Therefore, we tentatively plot $S_e(T)$ in Fig.~\ref{fig:Se} putting $S_e(0.16 $ K$)=0$ for each magnetic field; if $C_e/T \sim 0.047$ J/K$^2$mol at 0.16 K in zero field continues down to $T=0$, an error caused by this assumption would be $S_e(0.16 $ K$)-S_e(T=0) = 7 \times 10^{-3}$ J/K mol, which is negligibly small.
As a next step, the $S_e(T)$ curves for $\mu_0 H \neq 0 $ T are vertically shifted so that the Maxwell relation:
\begin{equation}
\label{eq:Maxwell} (\partial S_e/\partial (\mu_0H))_T=(\partial M/\partial T)_{\mu_0H}
\end{equation}  
is consistently satisfied at 5 K by both the $S_e$ and $M$ data shown in Fig.~\ref{fig:MT}.
The maximum shift of $S_e$ required for the adjustment is 0.02 J/K mol, which is invisible in Fig.~\ref{fig:Se}. 
Two $S_e(T)$ curves for adjacent magnetic fields in Fig.~\ref{fig:Se} cross at a temperature ($3 \sim 3.5$ K), which coincides with the maximum temperature of the corresponding $M/H$ vs $T$ curve shown in Fig.~\ref{fig:MT}, demonstrating the consistency of the present data with eq.~(\ref{eq:Maxwell}).
We also confirmed that results of magnetocaloric effect measurements are consistent with the $S_e$ data shown in Fig.~\ref{fig:Se}.

There are two possible interpretations for the Schottky-like anomaly appearing at $\sim 3$ K: (1) a CEF excitation superimposed on a moderately mass-enhanced quasiparticle excitation, or (2) a strongly energy-dependent quasiparticle excitation itself.
In the case (1), $\gamma$ is of the order of $10^{-1}$ J/K$^2$mol and $\Delta C/\gamma T_c$ is not far from the BCS value, while in the case (2), $\gamma(T)$ has a strong temperature dependence and $\Delta C/\gamma(T_c) T_c \simeq 0.1$ is quite smaller than the BCS value.

In the case (1), the maxima at $3 \sim 3.5$ K in the $M/H$-vs-$T$ curves as well as the field-sensitive behavior of the Schottky peak indicate that the CEF first excited state lying at $E_1/k_B \sim 8$ K is magnetic.
In the CEF model proposed by Bauer {\it et al.}~\cite{BauerPRB}, where the $\Gamma_3-\Gamma_5$ excitation leads to the Schottky peak in $C(T)$ at 3 K, an entropy of $R \ln 2$ associated with the $\Gamma_3$ ground state should be hidden below 0.16 K in zero field, if it is assumed that 4$f$ electrons are well localized.
Since applying magnetic field should cause a small but detectable splitting of the $\Gamma_3$ doublet (e.g. an energy splitting $\Delta E =1.5$ K in 4 T is calculated using the CEF parameters of $x=-0.72$ and $W=-5.44$ K from ref.~\cite{BauerPRB}), a new low-$T$ peak would appear in $C(T)$ and thereby the hidden entropy would be released, i.e., $S_e(0 $ T$)-S_e(4 $ T$) \simeq R \ln 2$ at 0.16 K.
Our $S_e$ data shown in Fig.~\ref{fig:Se} are clearly against this scenario, and we conclude that no anomalous entropy is concealed below 0.16 K in zero field.~\cite{QKL}

If a $\Gamma_1$ singlet is the CEF ground state, the first excited state should be a magnetic triplet $\Gamma_5$ ($\Gamma_4$ in $T_h$ notation~\cite{Ta00}).
Observed anisotropy in $M$, i.e., $M(H//\langle110\rangle) \simeq M(H//\langle111\rangle) = 1.110$ $\mu_B/{\rm Pr} > M(H//\langle100\rangle) = 1.095$ $\mu_B/{\rm Pr}$ at 1.9 K in 7 T, asserts the $\Gamma_1-\Gamma_5$ level scheme (see Fig. 7 of ref.~\cite{AoPrFe4P12PRB} for a calculation of $M$ for a single-site Pr ion), although the absence of clear downward curvature in $M(H)$ curves for all the three field directions (see Fig.~\ref{fig:MT} for $H//\langle100\rangle$) is not in agreement with the simple CEF predictions not only for the $\Gamma_1-\Gamma_5$ but also for the $\Gamma_3-\Gamma_5$ level schemes.
The $\Gamma_1-\Gamma_5$ level scheme is consistent with the fact that $S_e$ is lower than $R \ln 4$ at 8 K and increasing gradually.
The maximum value of $C$ at $\sim 3$ K is smaller than 8.51 J/K mol expected for the singlet-triplet Schottky peak,\cite{CEF12} indicating that the triplet excited state has a energy dispersion due to Pr-Pr magnetic interactions.

The FIOP appears in the field region where one level out of the $\Gamma_5$ triplet goes down due to the Zeeman effect and effectively degenerates with the ground state.
Actually, $S_{e}$ shown in Fig.~\ref{fig:Se} increases with increasing field below $\sim 3$ K and seems to show a tendency of saturation (plateau) at $R\ln2$ when the FIOP is fully developed in fields above 8 T.
Therefore the formation of the FIOP probably needs the degree of freedom possessed by the quasi-degenerate doublet formed in the high fields.
We speculate that the order parameter is of an antiferroquadrupole accompanied by a field-induced antiferromagnetism (AFM), as observed in CeB$_6$~\cite{FujitaCeB6,CeB6Sera2001} and TmTe~\cite{TmTe,TmTeLink1998}.
In these compounds, the quadrupole ordering temperature shifts to higher temperatures with increasing field.
In PrOs$_4$Sb$_{12}$, the negative Curie-Weiss temperature $\Theta_{\rm CW} = -6.6$ K determined below 20 K as shown in the inset of Fig.~\ref{fig:MT} indicates the existence of AFM correlations.
Therefore, by field-induced AFM components, the FIOP could be energetically stabilized leading to $d T_x/d H > 0$.
Note that preliminary data of $d \rho/d T$ indicate that $d T_x/d H$ changes to negative above 10 T~\cite{SuDHvA}, as similarly observed in both CeB$_6$ and TmTe.

In the case (2), the Schottky-like peak in $C_e(T)$, similar to the one observed at $\sim 7 $ K in CeNiSn~\cite{Izawa96}, implies the existence of a hybridization gap in the energy spectrum of the renormalized quasiparticle excitations in PrOs$_4$Sb$_{12}$~\cite{Transport}.
From the $C_e(T)$ data, the size of the gap is roughly estimated to be $\Delta_K/k_B \sim 8 $ K, which is four times larger than $T_c$.
This fact suggests that the superconductivity appears in the temperature region where the gap structure is well developed.
Figure~\ref{fig:CT} clearly shows that the gap structure is sensitive to magnetic field and is destroyed above $\sim 5$ T.
In these fields, the quasiparticle density of states at the Fermi energy increases and consequently developed RKKY-type AFM interactions would help to form the FIOP.

For a clear distinction between the cases (1) and (2), no decisive experimental facts are available at this stage.
However, this point should be clarified to understand the first $4f^2$-based HFSC in PrOs$_4$Sb$_{12}$.


We thank H. Kotegawa for informative discussions. 
This work is supported partly by a Grant-in-Aid for Scientific Research from the Ministry of Education, Science and Culture and by the REIMEI Research Resources of Japan Atomic Energy Research Institute.

{\it Note added.}- The effect of the deviation from the $T^{-2}$-dependence of the Pr nuclear contribution becomes nonnegligible in high fields ({\it e.g.}, see Fig. 1 in Ref. 23). If one follow the analysis described in Ref. 3, $m_{\rm Pr}=1.15 \mu_{\rm B}$/Pr is obtained in 8 T. Note that the long heat pulse ($\sim 1 $ min), the longtime $T$-response measurement ($> 10 $ min) and our data fitting procedure fully dealing with the tau-2 effect caused by long nuclear-spin-lattice-relaxation times allow us the precise measurements of the nuclear contribution.
After completion of this paper, we became aware of the result of Maple {\it et al.} who observed anomaly in $\rho$ corresponding to the FIOP.~\cite{MapleOrb}


\begin{references}
\bibitem{QKL}
We cannot deny a possibility that $4f$ electrons with a $\Gamma_3$ ground state are strongly hybridized with conduction electrons using quadrupole degrees of freedom, forming an effective singlet ground state in such a quadrupole Kondo lattice.
In this case, the entropy of $R \ln 2$ associated with the $\Gamma_3$ ground state will be released continuously from higher temperatures of the order of the quadrupole Kondo temperature.
\end{references}
\end{document}